\documentclass[%
 reprint,
 superscriptaddress,
 amsmath,amssymb,
 aps, pra,
 twocolumn,
]{revtex4-2}

\usepackage{bm}
\usepackage{graphicx}
\usepackage{amsmath}
\usepackage{appendix}
\usepackage{pythonhighlight}
\usepackage{enumitem}
\usepackage{braket}
\usepackage{multirow}
\usepackage{qcircuit}
\usepackage{soul}
\usepackage[caption = false]{subfig}
\usepackage{amsmath}
\usepackage{tabularx}
\usepackage{textgreek}
\usepackage{changes}
\usepackage{xcolor}
\usepackage[colorlinks=true,linkcolor=blue,anchorcolor=blue,citecolor=blue]{hyperref}
\DeclareMathOperator{\iSWAP}{\mathsf{iSWAP}}
\DeclareMathOperator{\CZ}{\mathsf{CZ}}
\DeclareMathOperator{\CNOT}{\mathsf{CNOT}}
\DeclareMathOperator{\CXSWAP}{\mathsf{CXSWAP}}
\DeclareMathOperator{\SWAP}{\mathsf{SWAP}}
\DeclareMathOperator{\DCNOT}{\mathsf{DCNOT}}

\begin{document}

\title{Routing-based technique for defect mitigation in quantum error correction}

\author{Runshi Zhou}
\affiliation{
 Department of Computer Science and Technology,
 Tsinghua University, Beijing, P.R.China}

\author{Fang Zhang}
\thanks{Corresponding authors: Fang Zhang (\href{mailto:zhangfang@iqubit.org}{zhangfang@iqubit.org}), Linghang Kong (\href{mailto:linghang@iqubit.org}{linghang@iqubit.org}), and Jianxin Chen (\href{mailto:chenjianxin@tsinghua.edu.cn}{chenjianxin@tsinghua.edu.cn})}
\author{Linghang Kong}
\thanks{Corresponding authors: Fang Zhang (\href{mailto:zhangfang@iqubit.org}{zhangfang@iqubit.org}), Linghang Kong (\href{mailto:linghang@iqubit.org}{linghang@iqubit.org}), and Jianxin Chen (\href{mailto:chenjianxin@tsinghua.edu.cn}{chenjianxin@tsinghua.edu.cn})}
\author{Feng Wu}
\author{Hui-Hai Zhao}
\affiliation{Zhongguancun Laboratory, Beijing, P.R.China}%

\author{Jianxin Chen}
\thanks{Corresponding authors: Fang Zhang (\href{mailto:zhangfang@iqubit.org}{zhangfang@iqubit.org}), Linghang Kong (\href{mailto:linghang@iqubit.org}{linghang@iqubit.org}), and Jianxin Chen (\href{mailto:chenjianxin@tsinghua.edu.cn}{chenjianxin@tsinghua.edu.cn})}
\affiliation{
 Department of Computer Science and Technology,
 Tsinghua University, Beijing, P.R.China}

\date{March 2, 2026}

\begin{abstract}
As quantum chips scale up for large-scale computation, hardware defects become inevitable and must be carefully addressed. In this work, we introduce Halma, a defect mitigation technique empowered by an expanded native gate set that incorporates the $\iSWAP$ gate alongside the conventional $\CNOT$ gate. Halma emerges as a supplementary technique within the defect mitigation toolbox, offering effective mitigation of ancilla qubit defects encountered during surface code stabilizer measurements while maintaining compatibility with existing superstabilizer-based methodologies. Halma introduces zero reduction in the spacelike distance of the code without further sacrifice to the timelike distance. Numerical simulation suggests that in comparison to previous methods, Halma could provide an order of magnitude improvement in the average logical error rate under realistic experimental settings, leading to a $\sim3\times$ reduction in the footprint of a teraquop. These results clearly demonstrate the capability of Halma in easing the near-term realization of fault-tolerant quantum computing on hardware with fabrication defects, and exemplifies how leveraging intrinsic hardware capabilities can enhance quantum hardware performance.

\end{abstract}

\maketitle


\section{Introduction}

Quantum error correction schemes have long relied on $\CNOT$ gates to generate and propagate parity information. Recent advancements, such as the demonstration of a surface code based on $\iSWAP$ gates~\cite{Dynamic_surface_codes2025}, highlight the feasibility of non-$\CNOT$-based quantum error correction.
The recently proposed AshN scheme reveals a significantly richer toolbox for superconducting qubits than previously anticipated, enabling the high-fidelity generation of arbitrary two-qubit gates~\cite{chen_one_2024,chen2025efficient}. Yet, the power unlocked by this expanded gate set has not been thoroughly studied in the context of quantum error correction.

Exploring what can be achieved with access to the new functionalities, we adopt a bottom-up approach and revisit the implementation of the surface code on superconducting qubits, focusing specifically on the issue of handling ancilla qubit defects. The surface code~\cite{bravyi_quantum_1998} is particularly favored, especially on superconducting platforms, due to its local requirement for qubit connectivity. However, since superconducting qubits are artificial atoms (engineered structures), the imperfect manufacturing process would lead to  1\%--2\% of defects, i.e., unusable qubits or qubit couplers, for a variety of causes~\cite{lisenfeld_electric_2019, smith_scaling_2022}. Because of the existence of defects, it is very unlikely that the fabricated hardware would satisfy the qubit connectivity required for surface code implementation, especially as the code size increases.

To mitigate the effect of fabrication defects, the current leading approach is based on a technique named superstabilizer~\cite{siegel_adaptive_2023, wei_low-overhead_2024, strikis_quantum_2023}, which excludes the defective region from syndrome extraction by instead measuring the larger superstabilizers encircling the defects, which are products of the original stabilizers in that region. Though superstabilizer-based approaches are well suited for handling defects on data qubits (which contain the encoded information), their performances are unsatisfactory when handling defects on ancilla qubits (which perform stabilizer measurements) and typically incur a significant reduction in the code distance and higher susceptibility to physical noise.

It is not surprising that leveraging more intrinsic hardware capabilities can enable advances beyond conventional approaches. By adapting an expanded, yet experimentally realistic native instruction set containing both the $\CZ$ gate and the $\iSWAP$ gate, we develop Halma, a routing-based technique that mitigates defects on ancilla qubits. During the syndrome extraction circuit of Halma handling such defects, the stabilizers measured near the defective region is the same as those of the defect-free surface code, thus the spacelike distance of the code is unchanged. Halma handles single ancilla qubit defect on a small-distance surface code with a logical error rate that is only $\sim1.5\times$ that of the defect-free code, under a realistic hardware error rate of $10^{-3}$. Compared to previous methods based on a single native two-qubit gate~\cite{siegel_adaptive_2023, wei_low-overhead_2024, strikis_quantum_2023}, Halma provides remarkable reductions in logical error rates on the scale of $\sim10\times$ for an average surface code with distance-11 and defects randomly distributed at a rate of 2\%, and the reduced logical error rate leads to an estimated teraquop footprint that is three times lower.

\section{Relevant background}
\subsection{QECC and surface code}

To detect and correct physical errors that occur during computation and ensure that they do not affect the logical information that is processed, quantum error correcting codes are employed to encode logical qubits into many physical qubits. A quantum error-correcting code is defined by its stabilizers, which are commuting operators that are constantly measured during syndrome extractions. Typically, ancilla qubits, each associated with a certain stabilizer, are used to collect information from the data qubits that are in the support of the stabilizer, so that they can be measured without disturbing the information encoded. The results of the stabilizer measurements are known as the syndrome of the code, which, after decoding, should ideally reflect the physical errors that have occurred in the code. 

\begin{figure}[ht]
\centering
\subfloat[3 $\times$ 3 surface code]{\includegraphics[width=  0.35\columnwidth]{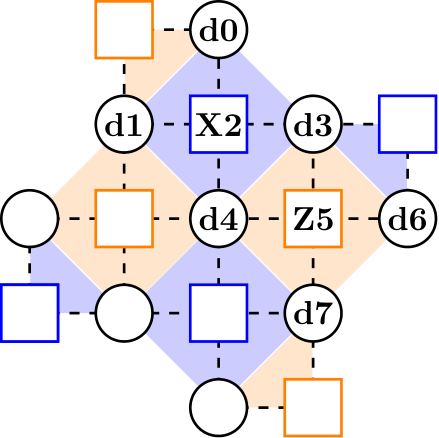}}\hspace{3mm}
\subfloat[Syndrome extraction circuit]{\includegraphics[width=  0.35\columnwidth]{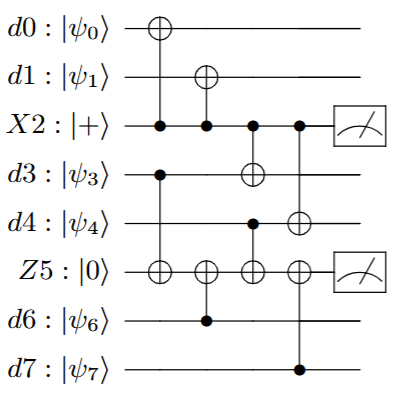}}
\caption{(a) A $3\times 3$ surface code for one round of syndrome extraction. Data qubits are illustrated as circles and X- and Z-ancilla qubits are illustrated as blue and orange squares, respectively. Qubit connectivity is labeled in dashed lines. (b) The circuit diagram for a local part of the syndrome extraction circuit.}
\label{surface_code}
\end{figure}

A distance $d$ surface code encodes 1 logical qubit in a $d\times d$ 2-dimensional array of data qubits and performs syndrome extraction using $d\times d-1$ ancilla qubits. A $3\times3$ surface code is shown in Fig.~\ref{surface_code}(a) as an example, where data qubits are illustrated as circles and X- and Z-ancilla qubits are illustrated as blue and orange squares, respectively. Each ancilla qubit is associated with a stabilizer, whose support contains the 4 (or 2 for stabilizers on the boundary) neighboring data qubits of that ancilla qubit. For example, the stabilizer associated with X2 is the product of Pauli-X operators on d0, d1, d3 and d4. This is a particularly favorable design because on typical hardware, such as superconducting chips, two-qubit gates can only be performed between nearest neighbors.

During a syndrome extraction circuit, each ancilla qubit performs two-qubit gates with its 4 (or 2) neighboring data qubits, to measure their associated stabilizer. In the most efficient depth-5 syndrome extraction circuit of a defect-free lattice, the two-qubit gates are performed in a certain order to ensure commutativity and reduce hook errors~\cite{dennis_topological_2002}. In this paper, we will define the order of the two-qubit gates using the ``Z-order'', which is the order of the data qubits with which a Z-ancilla qubit interacts, written in terms of their relative positions, with the top of the page as North. The X-order, i.e., the order for X-ancilla qubits, can be deduced from the Z-order, by flipping the second and third orientations. For example, a local part of the syndrome extraction circuit for the $3\times3$ surface code in Fig.~\ref{surface_code}, with a Z-order of \textit{NEWS}, which stands for North, East, West, South, is shown in Fig.~\ref{surface_code}(b).

One simple rule of thumb to ensure commutativity is that, on the two overlapping data qubits between an X- and a Z-stabilizer, the order by which the X- and Z-ancilla qubits interact with them must be the same. For example, between X2 and Z5, the two overlapping data qubits are d3 and d4. During a round of syndrome extraction, both d3 and d4 interact with Z5 before they interact with X2, so the circuit is legit. 

\subsection{Defects and superstabilizer}
Due to errors in fabrication, it is expected that a large quantum chip will almost inevitably contain some defects, such as qubit defects and coupler defects. A qubit defect happens when a physical qubit does not exist where it is supposed to, or has an abnormally higher physical error rate than other qubits. A coupler defect results in an abnormally high physical error rate when performing two-qubit gates via a certain coupler between two qubits, which makes the coupler effectively unusable. 

\begin{figure}[ht]
\centering
\subfloat[X-superstabilizer]{\includegraphics[width = 0.25\textwidth]{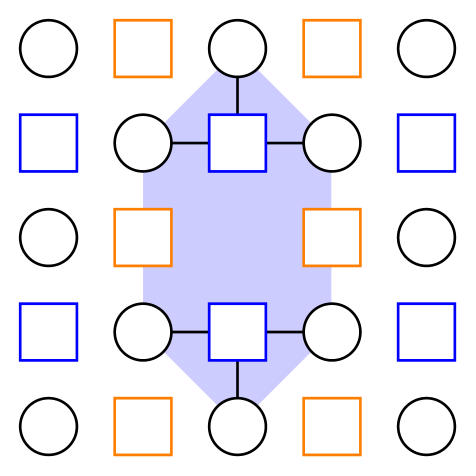}}\hspace{5mm} 
\subfloat[Z-superstabilizer]{\includegraphics[width = 0.25\textwidth]{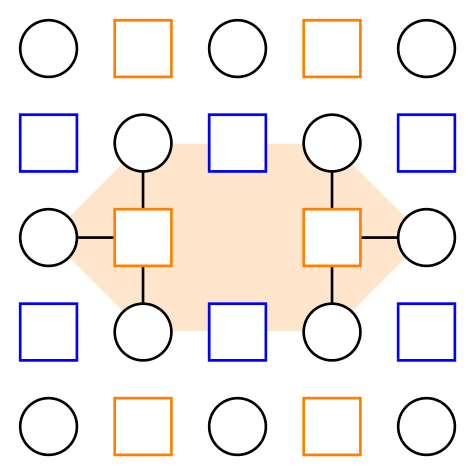}}
\caption{(a) The X-superstabilizer and (b) the Z-superstabilizer that are measured when a data qubit is defective. The 6 data qubits within the support of the X- and Z-superstabilizer are highlighted in the shaded region of the corresponding color.}
\label{SSforDQ}
\end{figure}

The current leading approach for performing error correction on defective lattices is to measure high-weight superstabilizers around the defective region on the lattice~\cite{siegel_adaptive_2023, wei_low-overhead_2024}. This approach performs very well when dealing with defective data qubits, as shown in Fig.~\ref{SSforDQ}. The value of the superstabilizer can be effectively obtained by measuring the associated gauge operators, which are the weight-3 operators around the defective data qubits that were used to be stabilizers of the surface code but are missing one data qubit in their supports due to the defect. Because the gauge operators do not all commute with each other, we need to alternately measure the X-superstabilizer and Z-superstabilizer between rounds of syndrome extraction.

\begin{figure}[ht]
\centering
\subfloat[X-superstabilizer]{\includegraphics[width = 0.25\textwidth]{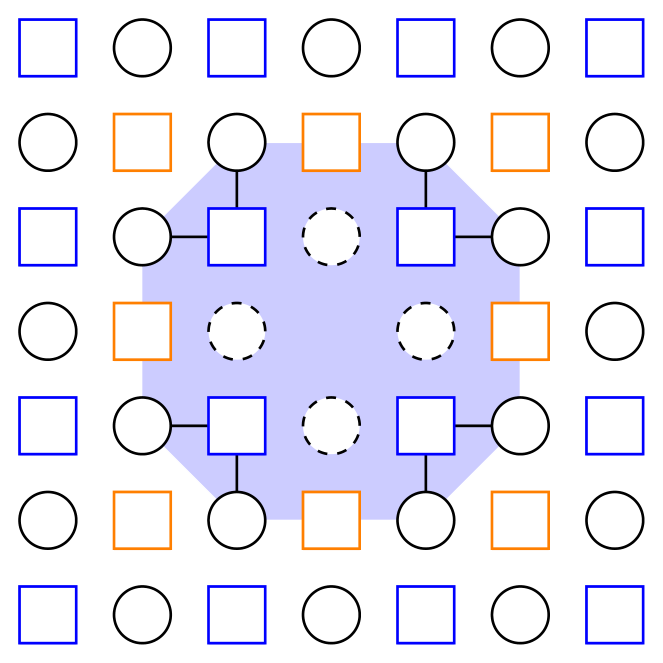}}\hspace{5mm} 
\subfloat[Z-superstabilizer]{\includegraphics[width = 0.25\textwidth]{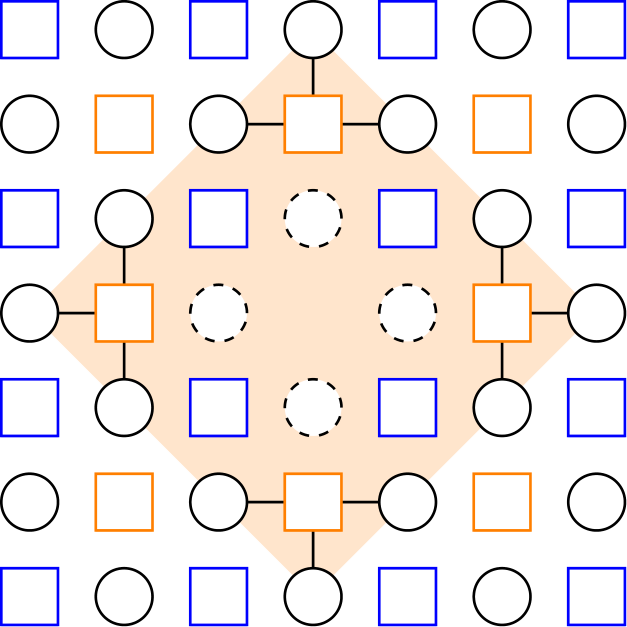}}
\caption{(a) The X-superstabilizer and (b) the Z-superstabilizer that are measured when an ancilla qubit is defective. The 4 dashed data qubits around the defective ancilla qubit are disabled, leading to a larger superstabilizer with a higher weight. }
\label{SSforCQ}
\end{figure}

The method of superstabilizer can also be applied to other types of defects, including defects on ancilla qubits and coupler defects. Those defects are usually handled by disabling some number of data qubits around the defective region so that the physical defective parts are never used. For example, a coupler defect is handled by disabling the data qubit which the defective coupler connects, whereas an ancilla qubit defect is handled by disabling the four neighboring data qubits as shown in Fig.~\ref{SSforCQ}. 

Unlike the data qubit defect cases, the superstabilizer approach to ancilla qubit defects is not very efficient, because it results in a significant qubit overhead and thus a reduction in the code distance. A single defect on an ancilla qubit will reduce the distance of the surface code by two, despite all the data qubits being physically intact. Additionally, the high weights of the superstabilizers also make their measurement very vulnerable to noise, leading to a higher logical error rate and lower threshold.

\subsection{Routing via CXSWAP Gates}

Addressing defects by qubit routing is not a new idea, and prior methods have managed defects by using $\SWAP$ gates to reassign qubits~\cite{nagayama_surface_2017}. However, these approaches often incur significant overhead, leading to longer and more complex syndrome extraction circuits and increased noise. They also often demand an algorithmic search to design the syndrome extraction circuit on the rest of the lattice.

In Halma, we assume that the quantum chip in use can natively perform both $\CNOT$ and $\CXSWAP$ gates while maintaining similar noise levels. A $\CXSWAP$ gate is a $\CNOT$ gate followed by a $\SWAP$ gate, first defined in~\cite{mcewen2023relaxing}. This is a reasonable assumption, particularly for superconducting platforms that support both XX and YY interactions, where the recent AshN gate scheme, as proposed by~\cite{chen_one_2024}, enables high-fidelity realizations of $\CZ$ and $\iSWAP$ gates simultaneously. Regarding the architecture of tunable coupling transmon, the $\CZ$ and $\iSWAP$ gates can be implemented on a same chip~\cite{Willow2024, Dynamic_surface_codes2025}. In this architecture, the ZZ interaction is activated by the dispersive XX+YY interaction that occurs between the coupler's $\left| 2 \right> $ level and data qubits' $\left| 11 \right> $ levels, which makes both $\CZ$ and $\iSWAP$ rely on the XX+YY interaction, allowing these two two-qubit gates to be implemented on the same qubits natively.

Since $\CNOT$ gates are equivalent to $\CZ$ gates up to single-qubit rotations, and $\CXSWAP$ gates are equivalent to $\iSWAP$, this approach leverages high-fidelity operations that are already achievable in current superconducting hardware~\cite{foxen_demonstrating_2020, krizan_quantum_2024}. More specifically, recent experimental work~\cite{chen2025efficient} demonstrates the implementation of arbitrary two-qubit gates---including $\CZ$ and $\iSWAP$---using the proposed AshN scheme. Notably, both gates are achieved with identical durations, leading to comparable fidelities when decoherence is the dominant noise source. In addition, ~\cite{sung2021realization} also demonstrates the realization of both $\CZ$ and $\iSWAP$ gates with fidelities of $99.76\pm 0.07\%$ and $99.87\pm 0.23\%$, respectively. Although this paper assumes that $\CZ$ and $\iSWAP$ gates have comparable fidelities (to simplify the noise model), we have investigated the effects of differing their fidelities in Fig.~\ref{noise_factor}. The simulation results from this analysis confirm that our core conclusion remains unchanged.

The additional $\CXSWAP$ gate in our instruction set essentially offers the option of performing a cost-free qubit routing whenever a $\CNOT$ gate is performed in the syndrome extraction circuit. This allows us to overcome the limited connectivity on a typical superconducting lattice, and measure the stabilizer associated with the defective ancilla qubit without lengthening the circuit or producing too much additional noise. 

The $\CXSWAP$ gate is also known as the double $\CNOT$ gate ($\DCNOT$), since it is logically equivalent to two $\CNOT$ gates acting in opposite directions. For logical clarity, we refer to this gate as $\CXSWAP$. We emphasize that it is a single two-qubit gate, and can be performed via $\iSWAP$ gate up to single-qubit rotations.

\begin{equation}
\Qcircuit@C=1em@R=1.5em{
&\targ      &\ctrl{1} &\qw & & &\ctrl{1}&\ctrl{1}&\targ      &\ctrl{1} &\qw & & &\ctrl{1}&\qswap&\qw\\
&\ctrl{-1}  &\targ    &\qw & \raisebox{2em}{=} & &\targ&\targ&\ctrl{-1}  &\targ    &\qw&\raisebox{2em}{=}& &\targ&\qswap\qwx&\qw
 }
\end{equation}

Note that the inverse operation of a $\CXSWAP$ gate is also a $\CXSWAP$ gate, but with the role switched between the target and control qubits. One can easily see that from the double $\CNOT$ interpretation, or by decomposing a $\CXSWAP$ gate into a $\CNOT$ gate followed by a $\SWAP$ gate. To reverse, it would require a $\SWAP$ gate followed by a $\CNOT$ gate, which is equivalent to a $\CNOT$ gate followed by a $\SWAP$ gate, but with the target qubit and control qubit reversed.

\begin{equation}
    \Qcircuit@C=1em@R=1.5em{
    &\ctrl{1}&\qswap&\qw & & &\qswap&\qswap&\ctrl{1}&\qswap&\qw& & &\qswap&\targ&\qw\\
    &\targ&\qswap\qwx&\qw & \raisebox{2em}{=} & &\qswap\qwx&\qswap\qwx&\targ&\qswap\qwx&\qw& \raisebox{2em}{=} & &\qswap\qwx&\ctrl{-1}&\qw
     }
\end{equation}

\section{Handling Single Ancilla Qubit Defect}

\begin{figure}[ht]
\centering
\includegraphics[width=0.5\columnwidth]{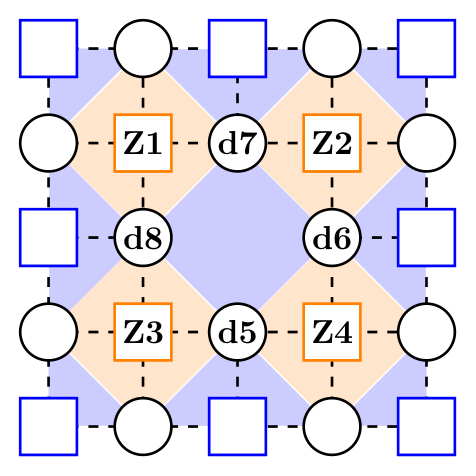} 
\caption{Labeling the qubits near an ancilla qubit defect}
\label{CheckDefectLabel}
\end{figure}

Without loss of generality, we begin with the assumption that we have a single defect on an X-ancilla qubit in the bulk of the code. The procedure can be generalized to Z-ancilla qubits by a horizontal flip along the defective qubit. For a clearer demonstration, we will label the other qubits in the surrounding area as shown in Fig.~\ref{CheckDefectLabel}.

\subsection{\textbf{W} round}
\begin{figure}[ht]
\centering
\subfloat[Circuit Diagram]{\includegraphics[width=0.4\textwidth]
{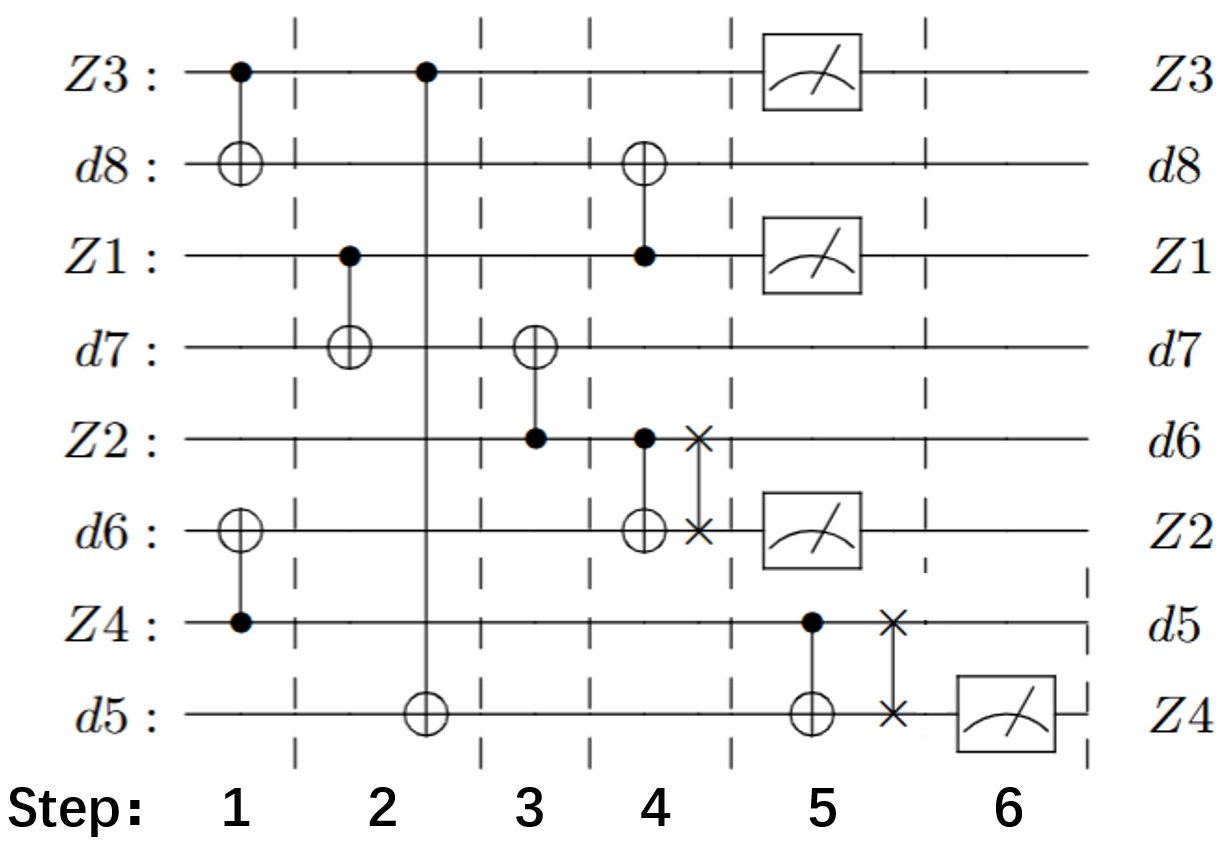}}\\
\subfloat[Step 1]{\includegraphics[width = 0.12\textwidth]{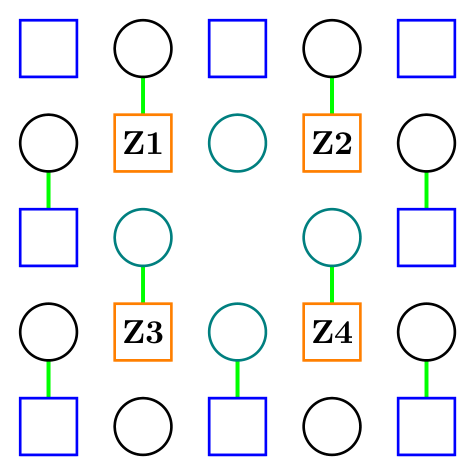}}\hspace{3mm} 
\subfloat[Step 2]{\includegraphics[width = 0.12\textwidth]{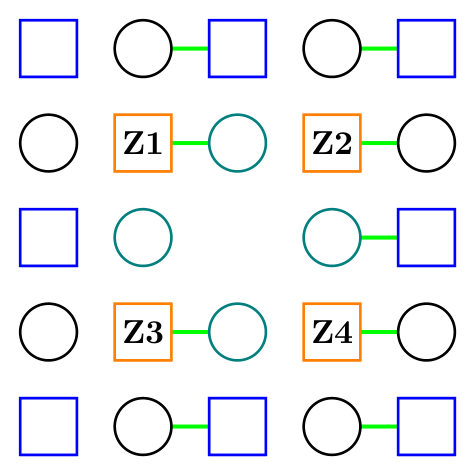}}\hspace{3mm} 
\subfloat[Step 3]{\includegraphics[width = 0.12\textwidth]{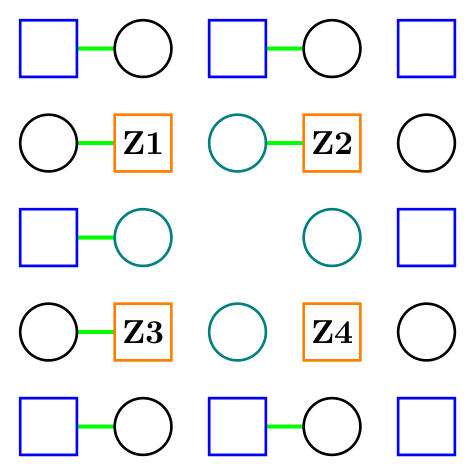}} \\
\subfloat[Step 4]{\includegraphics[width = 0.12\textwidth]{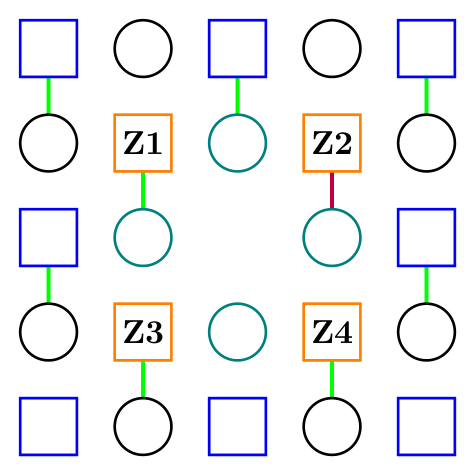}}\hspace{3mm} 
\subfloat[Step 5]{\includegraphics[width = 0.12\textwidth]{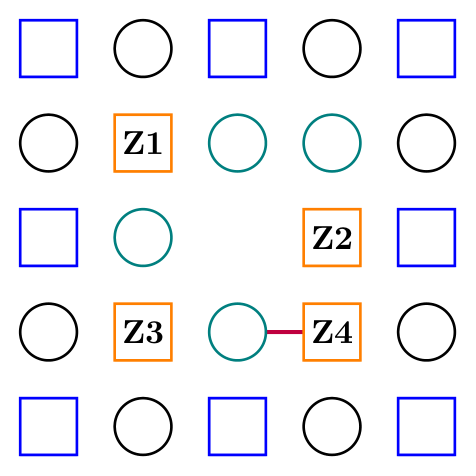}}\hspace{3mm} 
\subfloat[Step 6]{\includegraphics[width = 0.12\textwidth]
{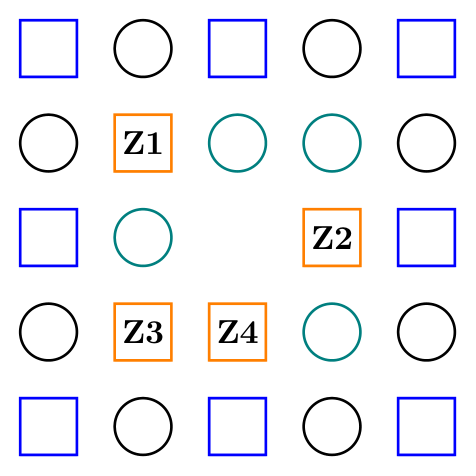}}\\
\caption{(a) The circuit diagram for the 6-step procedure of a \textbf{W} round in Halma. $\CXSWAP$ gates are denoted by a $\CNOT$ gate followed by a $\SWAP$ gate. (b-g) The layout illustration of the a \textbf{W} round. The data qubits within the support of the defective stabilizer are highlighted in teal. Green lines indicate $\CNOT$ gates and purple lines indicate $\CXSWAP$ gates. Each illustration correspond to one step of the procedure, and the qubit configurations are the configurations before the gates of that step are applied. Note that after a $\CXSWAP$ gate is applied, the circle and the square switch positions, as the associated qubits switch roles.}
\label{Even_Round}
\end{figure}

To handle a single ancilla qubit defect, we will first introduce two types of syndrome extraction, \textbf{W} rounds and \textbf{V} rounds. At \textbf{W} rounds, we measure all the stabilizers except the one associated with the defective ancilla qubit, while performing some routing to prepare for the following round, which is typically a \textbf{V} round. The circuit diagram with layout illustrations of a \textbf{W} round is presented in Fig.~\ref{Even_Round}.

A \textbf{W} round begins with the regular configuration of the surface code. In step 3, we postpone the $\CNOT$ gate between Z4 and d5 to step 5. This does not break the commutation, particularly because the stabilizer of the defective ancilla qubit is not measured. The depth of a \textbf{W} round alone is 6, but Z4 is idle in the first step of a \textbf{V} round, so step 6 can be absorbed into the beginning of a \textbf{V} round, and we can still integrate it into the depth-5 syndrome extraction circuit of surface code.

\subsection{\textbf{V} round}

\begin{figure}[ht]
\centering
\subfloat[Circuit Diagram]{\includegraphics[width = 0.4\textwidth]
{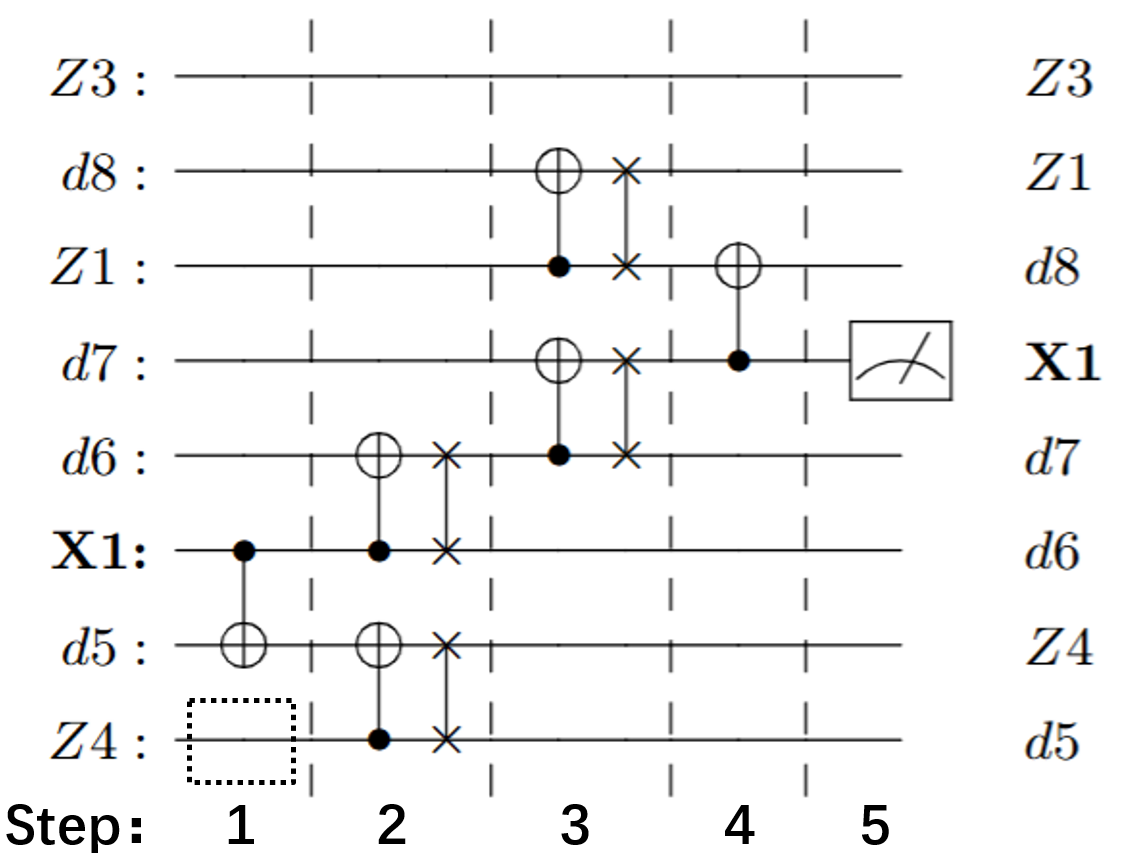}}\\
\subfloat[Step 1]{\includegraphics[width = 0.12\textwidth]{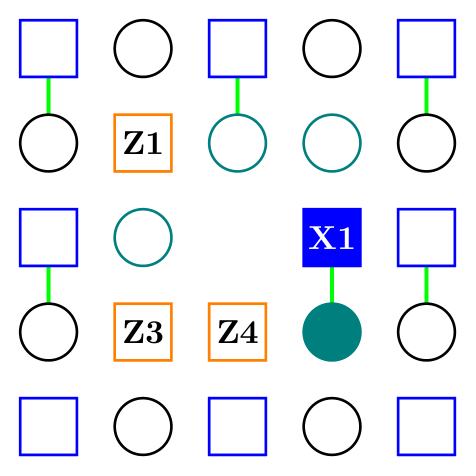}}\hspace{3mm} 
\subfloat[Step 2]{\includegraphics[width = 0.12\textwidth]{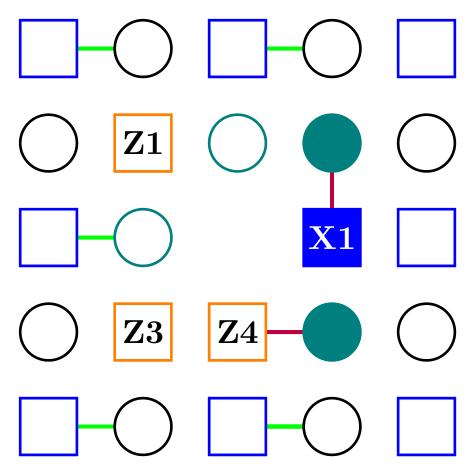}}\hspace{3mm} 
\subfloat[Step 3]{\includegraphics[width = 0.12\textwidth]{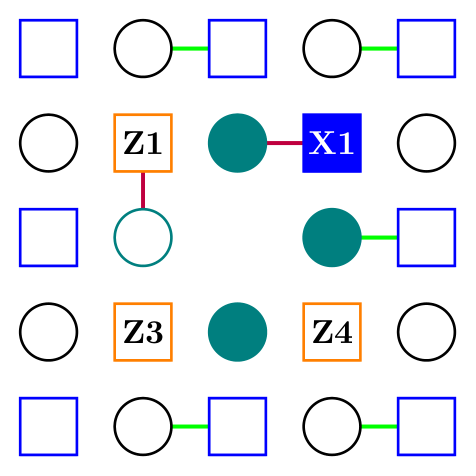}} \\
\subfloat[Step 4]{\includegraphics[width = 0.12\textwidth]
{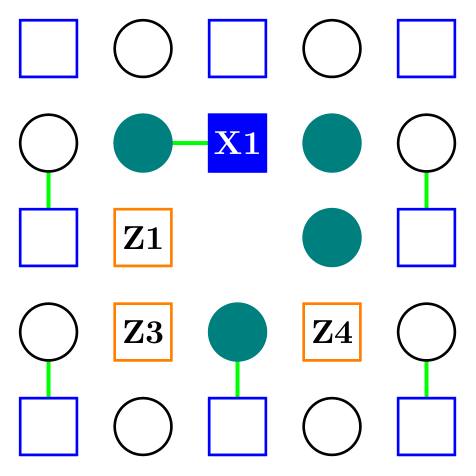}}\hspace{3mm} 
\subfloat[Step 5]{\includegraphics[width = 0.12\textwidth]{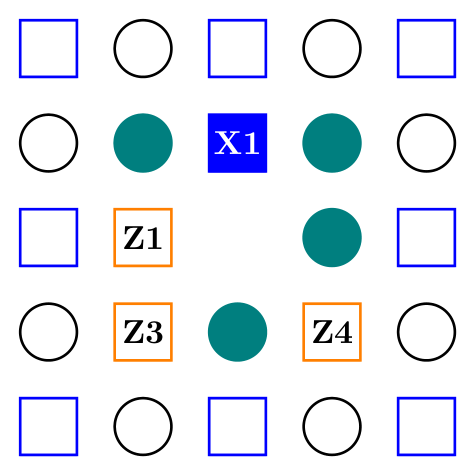}}\hspace{3mm} \\
\caption{(a) The circuit diagram for the 5-step procedure of an \textbf{V} round in Halma. The dashed box indicate when Z4 is idle in the beginning of the \textbf{V} round, where the extra time need for measuring Z4 in the end of the \textbf{W} round could be absorbed. (b-f) The layout illustration of the a \textbf{V} round. The data qubits within the support of the defective stabilizer are highlighted in teal, and we use filled circles to mark data qubits whose syndrome has been extracted by X1.}
\label{Odd_Round}
\end{figure}

At \textbf{V} rounds, we measure the defective stabilizer at the expense of disabling four surrounding stabilizer checks. One of the four surrounding ancilla qubits performs the stabilizer measurement that could not be done in \textbf{W} rounds, while the other three qubits help with routing. The diagrams for a \textbf{V} round are presented in Fig.~\ref{Odd_Round}.

A \textbf{V} round begins with the configurations at the end of a \textbf{W} round. The ancilla qubit Z2 now becomes an X-ancilla qubit, X1, which moves around and performs parity checks on the qubits in the support of the defective stabilizer, without affecting the other X-checks in the surrounding region. We note that in steps 2 and 3, the gates that we intend to perform on Z1 and Z4 are $\SWAP$ gates. Although we do not have $\SWAP$ gates in our instruction set, a $\CXSWAP$ gate with Z1 or Z4 as control is equivalent to a $\SWAP$ gate, since Z1 and Z4 are in the $\ket{0}$ state. This shortcut is also available for an X-ancilla qubit, simply by reversing the target and control of the gate.

\subsection{Between \textbf{W} and \textbf{V} rounds}
\begin{figure}[htbp]
\centering
\includegraphics[width=  0.8
\columnwidth]{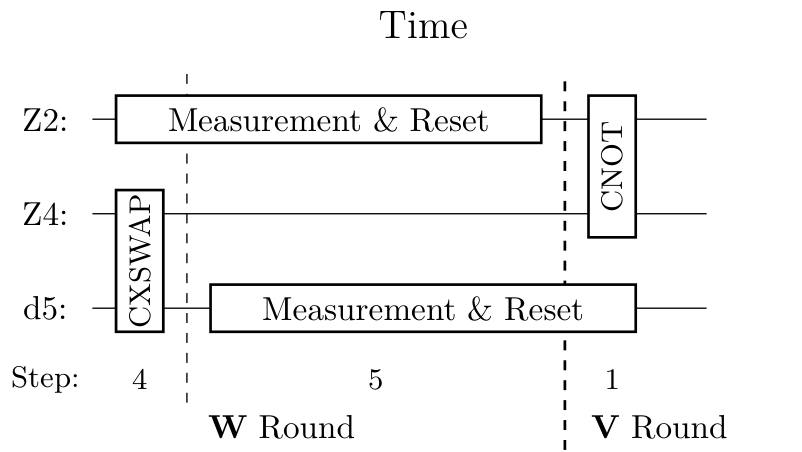} 
\caption{An illustration for the operations performed on qubit Z4 and d5 at the end of a \textbf{W} round and the beginning of a \textbf{V} round.}
\label{shift_measurment}
\end{figure}

At the end of a \textbf{W} round, due to the postponed 2-qubit gate between Z4 and d5, the Measure \& Reset operation on Z4 has to begin at a later time than the other ancilla qubits. However, this does not affect the overall circuit depth of Halma, because the additional time needed to finish the Measure \& Reset operation on Z4 can be absorbed at the beginning of the following \textbf{V} round, as illustrated in Fig.~\ref{shift_measurment}. Similar procedures as such have been used in other simulations such as \cite{O_Brien_2017} for superconducting platforms. Therefore, Halma can be integrated into the regular depth-5 syndrome extraction circuit.

In the simulation, this procedure is treated as three layers of operations, errors with non-idling operations are implied:
\begin{enumerate}
    \item 2-qubit gate between d5 and Z4; no \textit{idler errors} on the other qubits.
    \item Measurement \& Reset on all ancilla qubits; corresponding \textit{resonator idler errors} on the other qubits.
    \item 2-qubit gates in step 1 of the \textbf{V} round, \textit{idler errors} on the other qubits (including d5 and Z4 if they are idling).
\end{enumerate}
In this way, we are slightly overcounting errors on qubits Z4 and d5.

\subsection{WV\texorpdfstring{$\Lambda$}{Lambda}M Cycle.}
We have introduced the procedure of \textbf{W} round and the \textbf{V} round. There are three other types of rounds that are not introduced:
\begin{enumerate}
    \item An \textbf{M} round is a \textbf{W} round performed in reverse order, with measurements and reset at the very end.
    \item A $\bm{\Lambda}$ round is a \textbf{V} round performed in reverse order, with measurements and reset at the very end.
    \item A \textbf{III} round is similar to a regular round of a syndrome extraction circuit on a defect-free lattice, except that all the operations of the defective ancilla qubit are ignored.
\end{enumerate}

In \textbf{W}, \textbf{M} and \textbf{III} rounds, the stabilizer associated with the defective ancilla qubit is not measured. In \textbf{V} and $\bm{\Lambda}$ rounds, the stabilizer associated with the defective ancilla qubit is measured at the expense of temporarily disabling the surrounding ancilla qubits. Given the qubit configurations at the beginning and end of each type of round, we have the following rules. 
\begin{enumerate}
    \item To absorb the postponed operations as shown in Fig.~\ref{shift_measurment}, a \textbf{W} round, need to be followed by a \textbf{V} round. Likewise, an \textbf{M} need to be preceded by a \textbf{\textLambda} round.
    
    \item A \textbf{V} round must be followed by a \textbf{\textLambda} round.
    
    \item After a \textbf{\textLambda} round, one can perform either an \textbf{M} round or a \textbf{V} round.

    \item After an \textbf{M} round, the qubits are returned to their conventional configurations, thus it can be followed by either a \textbf{W} round or a \textbf{III} round.
    
    \item A \textbf{III} round can be followed by itself or by a \textbf{W} round.
\end{enumerate}

\begin{figure}[htbp]
\centering
\includegraphics[width=  0.5
\columnwidth]{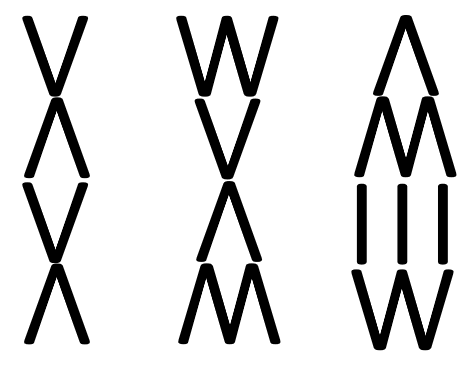} 
\caption{Three examples of round orderings, reading from top to bottom.}
\label{WVAM}
\end{figure}

These rules have a correspondence to the shape of the characters, as shown in Fig.~\ref{WVAM}. One exception is that we cannot perform an \textbf{M} round after a \textbf{W} round, but such a type of transition does not seem necessary, as we can convert both rounds to \textbf{III} rounds while measuring the same stabilizers.

Halma typically mitigates an ancilla qubit defect by a 4-round cycle of \textbf{WV\textLambda M}. During each 4-round cycle, each of the 5 stabilizers are measured twice, resulting in a timelike distance that is one-half of the defect-free code~\cite{gidney_stability_2022}. Meanwhile, since the stabilizers measured are the same as that of the defect-free surface code, the spacelike distance is not reduced.

\subsection{Defect on Boundary}
\begin{figure}[ht]
\centering
\subfloat[\textbf{W} Round]{\includegraphics[width = 0.3\textwidth]{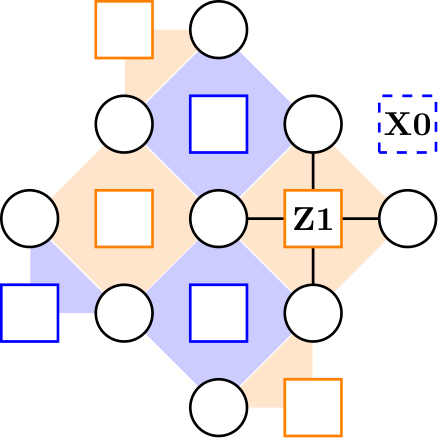}}\hspace{5mm} 
\subfloat[\textbf{V} Round]{\includegraphics[width = 0.3\textwidth]{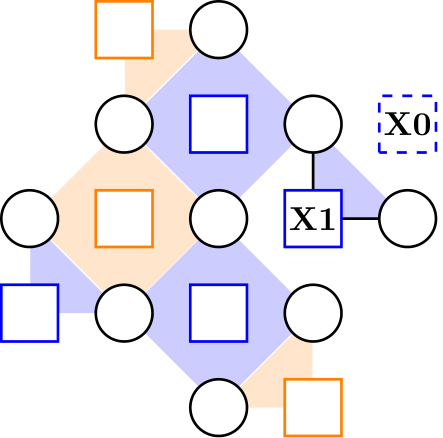}}
\caption{(a) A \textbf{W} round and (b) a \textbf{V} round handling an ancilla qubit defect on the boundary of a distance-3 surface code. The ancilla qubit defect happens at qubit X0. In \textbf{V} rounds, the adjacent ancilla qubit Z1 become X-ancilla qubit X1, and perform the stabilizer measurement for X0.}
\label{Boundary}
\end{figure}

Halma can handle defects on the boundary of a code in a similar way. The procedure is simpler because it does not involve any routing. The stabilizer of an ancilla qubit on the boundary only has two data qubits in its support, both of which are directly connected to another adjacent ancilla qubit. In a \textbf{V} round, we simply borrow that adjacent ancilla qubit to perform the parity measurement for the defective ancilla qubit, as shown in Fig.~\ref{Boundary}.

\subsection{Benchmarking Single Ancilla Qubit Defect.}
We examined the performance of Halma on a single ancilla qubit defect, located at the center of surface codes with distance $d = 4, 6, 8$, and compared it with other approaches.

\begin{figure}[h]
\centering
\subfloat[Superstabilizer]{\includegraphics[width=  0.6\columnwidth, trim={0 0 40 40}, clip]{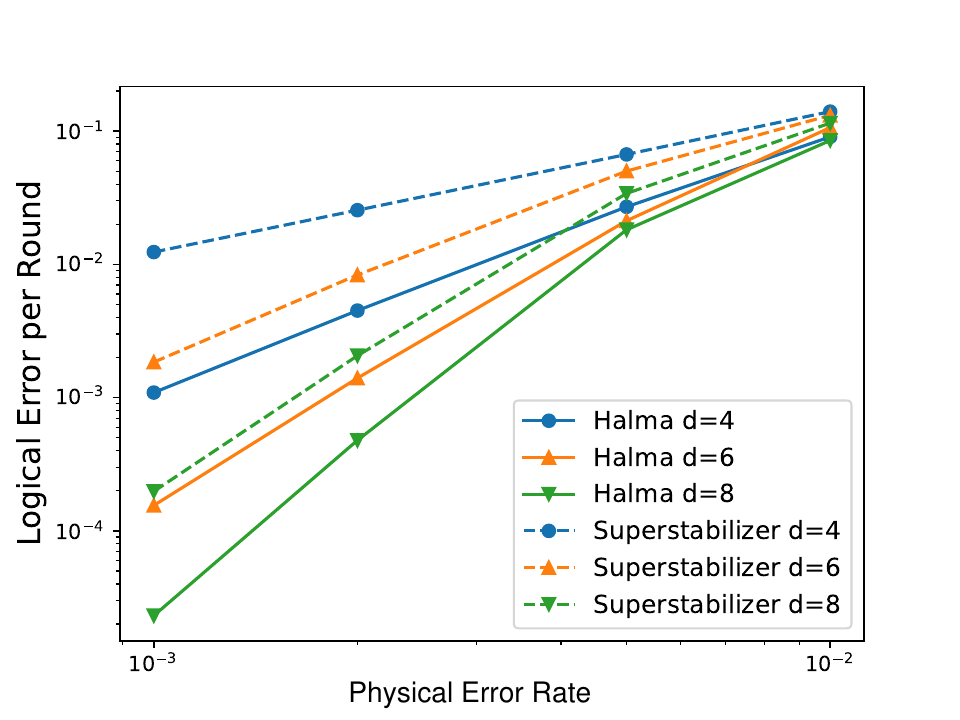} }\\
\subfloat[LUCI (Memory)]{\includegraphics[width=  0.5\columnwidth, trim={0 0 40 40}, clip]{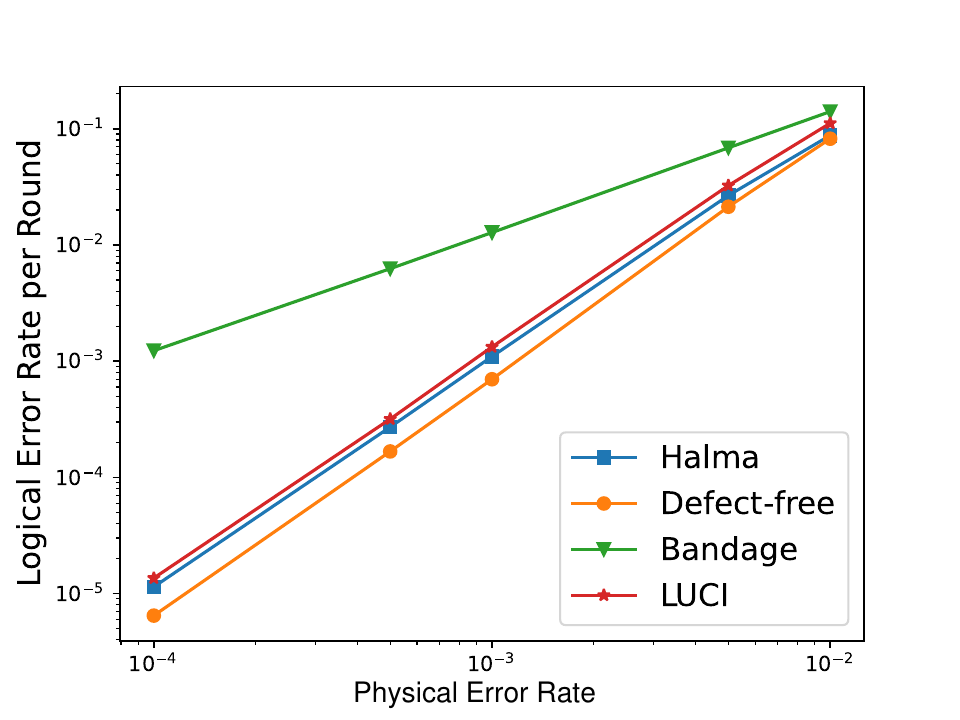}}
\subfloat[LUCI (Stability)]{\includegraphics[width=  0.5\columnwidth, trim={0 0 40 40}, clip]{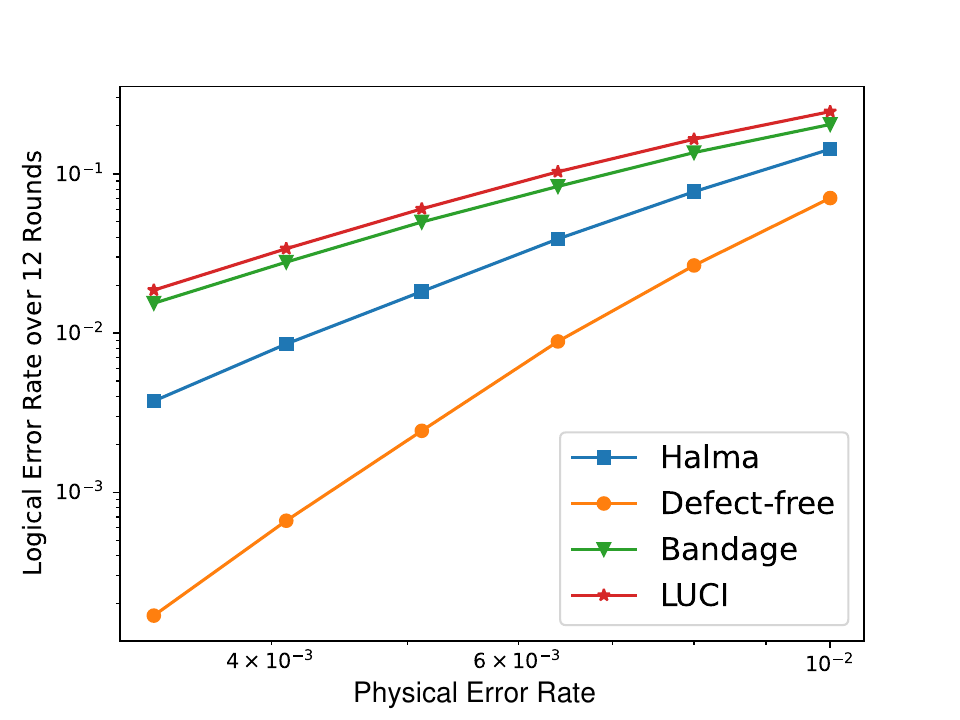}}
\caption{(a) The logical error rate of surface code with one ancilla qubit defect, handled by Halma (solid lines) and superstabilizer (dashed lines). Logical error rates are averaged between X and Z basis memories. The logical error rate of the $4\times 4$ surface code with a single defect in the (b) memory and (c) stability experiment. The conventional method is referred to as ``Bandage'' to avoid confusion, since superstabilizers are also used in LUCI.}
\label{halma_vs_supstab}
\end{figure}

As shown in Fig.~\ref{halma_vs_supstab}(a), our simulation shows a significant improvement in terms of the logical error rate per round compared to the standard superstabilizer approach. 

We also used manually written circuits on distance-4 surface code with defects, following the procedure of ~\cite{debroy_luci_2024}, to compare Halma with a more recent approach LUCI. We see from Fig.~\ref{halma_vs_supstab}(b,c) that Halma and LUCI have comparable performances in terms of logical memory, which is reasonable given that both approaches do not reduce the spacelike distance of the code. However, Halma has significantly better performance in a stability experiment in which 12 rounds of stabilizer measurements are performed~\cite{gidney_stability_2022}. This agrees with the fact that the timelike distance of Halma is twice as much as LUCI. This will make Halma more advantageous in performing logical operations on surface code with defects.

\section{Managing General Defect Patterns}
The simple model of surface code with a single ancilla qubit defect located at its center is certainly unrealistic. In this section, we will consider cases where there are multiple defects randomly distributed on different parts of a chip. We found that Halma is well compatible with the superstabilizer approach such that for most defective chips, we can handle ancilla qubit defects with Halma and handle data qubit defects with superstabilizers. Additionally, since the procedure of Halma is local in space, it can handle multiple defects while imposing a loose requirement on the sparsity of defects.

\subsection{Compatibility with Superstabilizer}
Since Halma only handles ancilla qubit defects, other approaches are required to resolve data-qubit defects and coupler defects, so that they compile to a complete workflow that handles arbitrary defect distribution in realistic lattices. Halma is well compatible with the superstabilizer approach as long as all the gauge operators in a superstabilizer can be measured within the same round. The most straightforward way to solve this situation is to alternate between global Pauli-X- and Pauli-Z-rounds. In an X(Z)-round, we perform a \textbf{W}, \textbf{M} or \textbf{III} round for all the Z(X)-ancilla qubit defects, and a \textbf{V} or $\bm{\Lambda}$ round for all the X(Z)-ancilla qubit defects, so that all the X(Z)-ancilla qubits are measured, whether they correspond to stabilizers or gauge operators.

\subsection{Defect Cluster}
\begin{figure}[ht]
\centering
\includegraphics[width=  0.5\columnwidth]{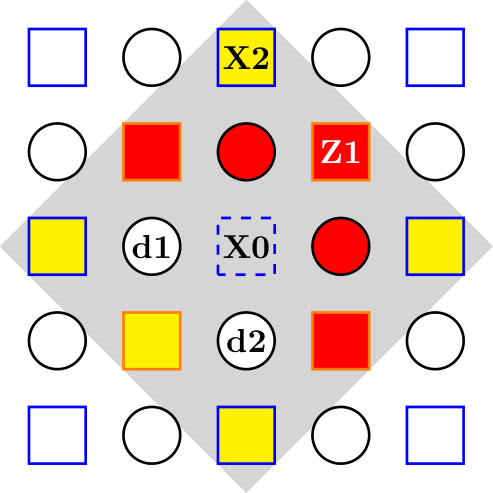} 
\caption{Labeling defect that will make it impossible to perform Halma on X0 (red), or where Halma cannot be simultaneously performed with X0 (yellow).}
\label{ClusterMap}
\end{figure}

Another potential problem is that when multiple defects cluster in a small region, the routing path of a certain ancilla qubit defect might be blocked, or conflict with that of a neighboring defect. Following the procedure introduced in Sec.~III, where there is an ancilla qubit defect at X0, we see that an additional defect on almost any other qubit in the shaded region of Fig.~\ref{ClusterMap} will affect the procedure of Halma handling X0. Fortunately, there are several methods that allow Halma to manage such defect clusters.

\subsubsection{Locally Modifying the Circuit}

In some cases, a defect cluster can be handled by locally modifying some part of the syndrome extraction circuit. For example, if data qubit d1(d2) in Fig.~\ref{ClusterMap} is also defective in addition to X0, then in \textbf{V} and $\bm{\Lambda}$ rounds, X1, the ancilla substituting X0, will need to measure a weight-3 gauge operator excluding d1(d2). Since the operations associated with d1(d2) in the original procedure of Halma do not contain any qubit routing, we could neglect any operations associated with it during the procedure of Halma, and the gauge operator would be successfully measured in every \textbf{V} and $\bm{\Lambda}$ rounds.

\subsubsection{Changing the Global Order}

When introducing the procedure of Halma previously, we assumed a certain global Z-order of the syndrome extraction circuit, namely, we begin with a \textbf{W} round with Z-order \textit{NEWS}, which fixes the Z-orders of the following 3 rounds. However, this choice is mostly arbitrary, as we could also choose other Z-orders that are 90-degree rotations, or reflections from it, though some Z-orders are preferred due to hook errors~\cite{dennis_topological_2002}.

\begin{figure}[ht]
\centering
\subfloat[NEWS]{\includegraphics[width = 3.5cm]{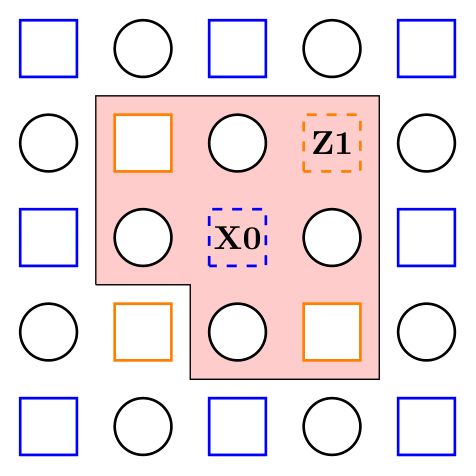}}\hspace{3mm}
\subfloat[SWEN]{\includegraphics[width = 3.5cm]{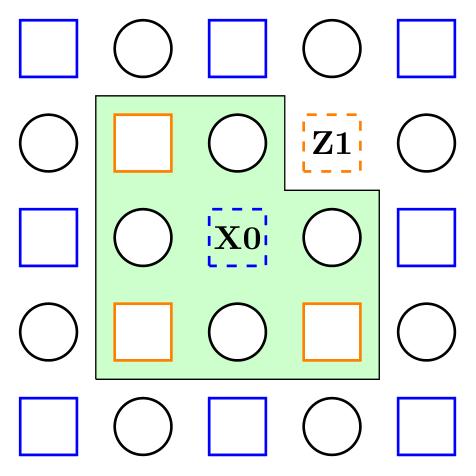}}
\caption{A diagram showing that by choosing a different Z-order, Halma can be made applicable on a defect cluster. The boxed region shows qubits that are involved in the procedure of Halma with a 4-round cycle beginning with a Z-order of (a) NEWS and (b) SWEN.}
\label{good_bad_cluster}
\end{figure}

For the procedure of Halma described in Sec.~III, there are some qubits near X0 where a second defect will make it impossible to perform Halma on X0. These qubits are labeled in red in Fig.~\ref{ClusterMap}. Nevertheless, if there is only one defect near X0, we can always find a Z-order where performing Halma on X0 is possible. For example, as shown in Fig.~\ref{good_bad_cluster} if we have defects on both X0 and Z1, in a 4-round cycle beginning with a Z-order of \textit{NEWS}, Halma cannot be performed on X0. But if the frame is rotated by 180 degrees, then Halma can be performed on X0 in a \textbf{WV\textLambda M} cycle beginning with a Z-order of \textit{SWEN}. Similarly, Halma can be performed on Z1 in a \textbf{WV\textLambda M} cycle beginning with a Z-order of \textit{ESNW}.

However, handling different defects in clusters on the same hardware might require \textbf{WV\textLambda M} cycle beginning with different global Z-orders. We typically prefer to change the Z-order at the end of an \textbf{M} or \textbf{III} round, where the qubits near a defect have a symmetric configuration, allowing us to perform the next \textbf{W} round with any Z-order of our choice. Changing the Z-order in the middle of a \textbf{WV\textLambda M} cycle is also feasible, but might require some routing overhead to adjust the qubits to the right configuration.

\subsubsection{Alternately Handle Neighboring Ancilla Qubit Defects}
There are also cases where Halma can be performed on individual ancilla qubit defects, but cannot be simultaneously performed on two defects in a cluster because their routing paths conflict. Those positions are colored yellow in Fig.~\ref{ClusterMap}. A similar situation also appears when certain defects within the same code can only be handled by Halma under different global Z-orders. In these cases, one simple resolution is to alternate between performing Halma on each of these defects, or sets of defects. For example, if there are defects on both X0 and X2, in the first 4-round cycle, Halma is performed only on X0, and it will be four \textbf{III} rounds for X2. Halma can then be performed only on X2 in the next cycle, and some overhead is required between the two cycles for reconfiguration. Note that applying this method will result in a further reduction in the timelike distance of the code. 

\subsubsection{Clusters on Boundaries}
The circumstance with handling ancilla qubit defects on the code's boundary is simpler, but imposes a stricter condition. In \textbf{V} and $\bm{\Lambda}$ rounds, a particular adjacent ancilla qubit measures the weight-2 stabilizer for the defective ancilla qubit on the code boundary. So, the only requirement is that this adjacent ancilla qubit is not defective. In this case, a two-qubit defect cluster can make it impossible to perform Halma, and we will have to resort to deforming the boundary of the code following \cite{wei_low-overhead_2024}.

\subsubsection{Robustness against Defect Clusters}
Using the three aforementioned methods, one can conclude that in the bulk region of the code, it would take at least two additional defects near an existing ancilla qubit defect to make it completely impossible to handle this defect with Halma, which imposes a loose requirement on defect sparsity and makes Halma applicable to most defective hardware.

By evaluating ancilla qubit defects on randomly generated defective lattices, we found that when the defect rate is 2\%, Halma can be directly applied to around 90\% of ancilla qubit defects without using the three methods previously discussed. With the assistance of these methods, Halma can handle up to around 99\% of ancilla qubit defects. Those two values increase further to 94\% and 99.7\% when the defect rate is reduced to 1\%.

It is worth mentioning that there is another type of defect cluster that Halma cannot handle swiftly. For a superstabilizer, if more than one of the ancilla qubits associated with its gauge operators is defective, we require that Halma can measure these gauge operators within the same rounds of a certain Z-order. This is sometimes impossible, or very difficult, due to other defects nearby. A simple example is two ancilla qubit defects located on the opposite sides of a data qubit defect. For cases like this to happen, it is straightforward to see that the cluster still needs to contain at least 3 defects. So, the previous conclusion holds nevertheless.

\subsection{Preliminary Studies on Defect-Handling Strategy}
Though the flexibility of Halma makes it adaptable to defect clusters, it also presents many choices that need to be made. First, for each cluster, we need to decide if the ancilla qubit defects within the cluster were to be handled by Halma or other compatible methods, such as the conventional superstabilizer approach of disabling the surrounding data qubits. Then, before each round of syndrome extraction of the code, we need to choose the global Z-order of this round, and the type of round (\textbf{W}, \textbf{V}, \textbf{\textLambda}, \textbf{M} or \textbf{III}) to perform on each ancilla qubit defect, while ensuring the transition between rounds is legal, or if it turns out preferable, forcing a transition by performing some manual routing. Additionally, we might choose to forbid certain Z-orders to prevent hook errors; to repeatedly measure higher-weight superstabilizers as discussed in~\cite{strikis_quantum_2023}; or to avoid further reducing the timelike distance by forsaking alternating between handling different ancilla qubit defects. An instruction of the aforementioned choices made is defined as a strategy, and a strategy module should produce strategies for any given defect distribution on a lattice.

One could build a strategy module based on the XXZZ 4-round cycle that was previously introduced when discussing the compatibility between Halma and superstabilizers, though strategies produced by this module could be suboptimal, as the global regulation imposed by the X- and Z-rounds is stricter than necessary.

\begin{figure}[ht]
\centering
\subfloat[Cluster Pattern]{\includegraphics[width=  0.25\columnwidth]{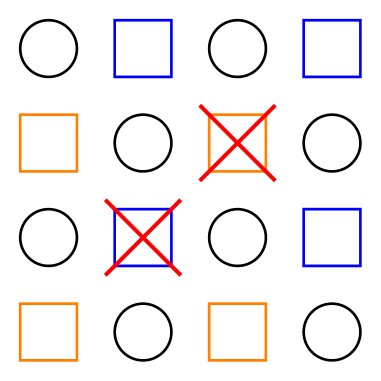}}\hspace{5mm}
\subfloat[Logical Error Rate]{\includegraphics[width=  0.35\columnwidth]{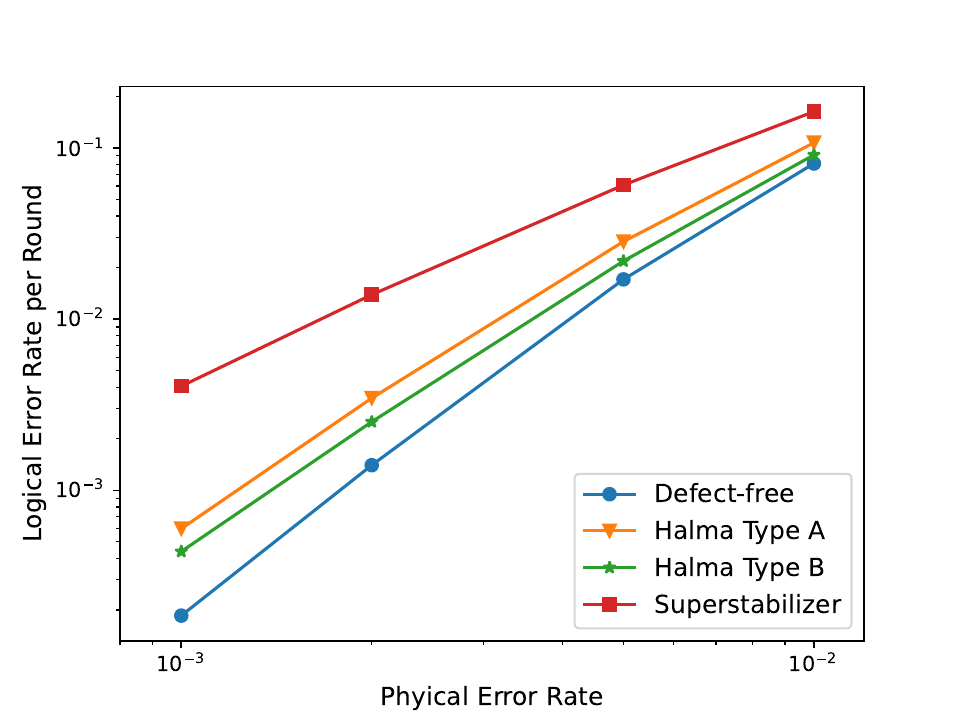}}
\caption{(a) An illustration of the defect cluster.(b) The simulated logical error rate per round of the defect cluster handled by either Halma or superstabilizer. Type A and Type B differ in which of the two defects is handled by Halma in the first 4 rounds.}
\label{2cluster}
\end{figure}

Finding and implementing a good strategy module would require sophisticated algorithms and complicated circuitry. We leave the study of this topic to future research, as our purpose for now is only to showcase the power of Halma. Nevertheless, we are optimistic that even using the suboptimal strategies, Halma will still outperform conventional superstabilizer approaches in terms of logical error rates. We performed a preliminary simulation estimating the performance of Halma on defect clusters. We focus specifically on a case with a defect cluster illustrated in Fig.~\ref{2cluster}(a) in the bulk of a distance-5 surface code. The resulting circuits have a timelike distance that is 1/4 of the defect-free code. The simulated logical error rates are plotted in Fig.~\ref{2cluster}(b), where we see Halma still have significantly lower logical error rates in comparison to the conventional superstabilizer method.

\subsection{Benchmarking General Defect Patterns}

\begin{figure}[ht]
\centering
\includegraphics[width=  0.6\columnwidth]{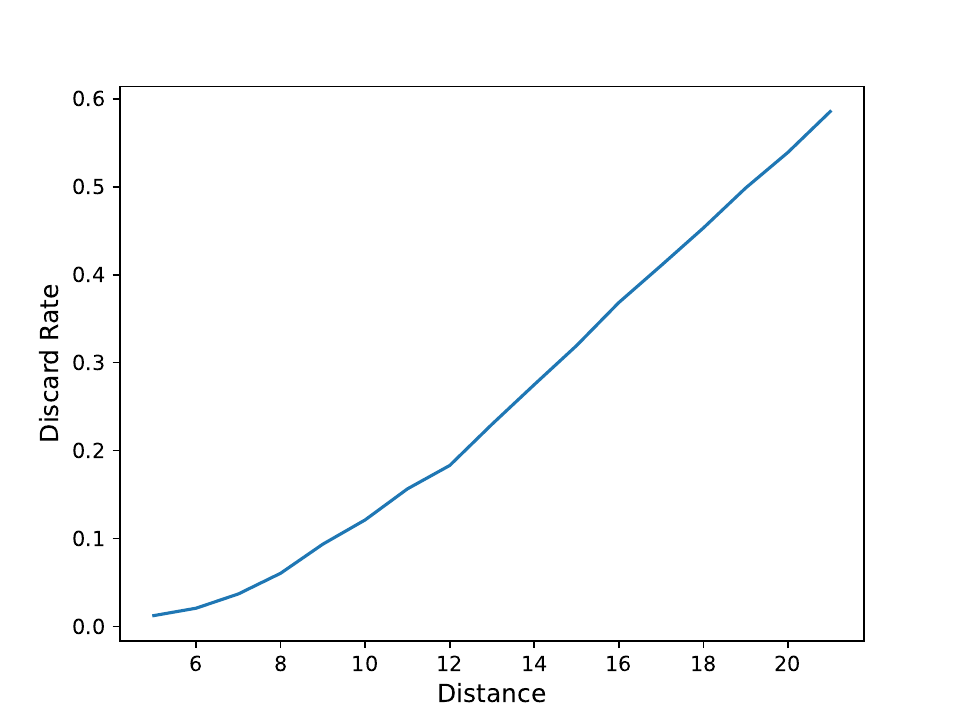}
\caption{The discarding rate simulated using 10,000 samples for each distance at the condition of this simulation.}
\label{discard_rate}
\end{figure}

To examine the performance of Halma on realistic chips, we generated 1000 defective surface code lattices with different distances, where defects are randomly distributed at a rate of 2\%. Each defective lattice is treated with either a superstabilizer approach, following the procedure of~\cite{wei_low-overhead_2024}, or the Halma approach introduced in this paper. We found that Halma leads to significant improvements in the logical error rate and the teraquop footprint. The procedure of our simulation is as follows:

{\parindent0pt
\vspace{2mm}
\textbf{Distribute Defects:}

First, we generate a defect-free surface code lattice with distance $d=5,7,9,11,13,15,21$. We randomly assign if each qubit on that lattice is defective or functional, with a defect rate of $r=0.02$.

\vspace{2mm}
\textbf{Deform Boundary:}

Then, we handle the defects on the boundary of the code by disabling certain boundary qubits. Essentially, we want to ensure that, after deformation, a data qubit on the X-boundary is within the support of two functional X-stabilizers and one functional Z-stabilizer. A similar rule applies to data qubits on the Z-boundary, and for data qubits that are on both the X- and Z-boundaries, they need to be contained within one functional stabilizer of each basis. A more detailed procedure is discussed in~\cite{wei_low-overhead_2024}, which is used for the superstabilizer group. For Halma, we deform the boundary in a similar procedure, except that if a defective ancilla qubit on the lattice boundary can be handled by Halma, we consider its associated stabilizer to be functional during deformation.

\vspace{2mm}
\textbf{Generate Strategy:}

The strategy for superstabilizers is straightforward, where we handle every data qubit defect and ancilla qubit defect with superstabilizers. For Halma, we employ a baseline strategy in this simulation. While handling all data qubit defects with superstabilizers, we examine every ancilla qubit defect and see if there is a second defect within its vicinity. If there is a second defect, we handle it with the superstabilizer approach, by disabling the 4 neighboring data qubits. If there is no second defect, we handle it with Halma. This is a baseline strategy that is only meant to simulate the minimum improvement that Halma can provide. 

\vspace{2mm}
\textbf{Post-Selection:}

We also estimated the maximum potential improvement of Halma by making a separate post-selected set of samples. Essentially, after the \textbf{Generate Strategy} step, we implement an additional step, checking if all ancilla qubit defects are handled by Halma. If not, we discard the case, return to the very beginning, and regenerate a new defect distribution. This post-selection condition ensures that all the ancilla qubit defects can be handled by Halma without conflict with other nearby defects. The percentages of discarded lattices for every distance are estimated and displayed in Fig.~\ref{discard_rate}.

\vspace{2mm}
\textbf{Generate Circuit:}

Afterward, we generate a $d$-round syndrome extraction circuit in Stim~\cite{gidney_stim_2021} based on the given defect distribution and handling strategy. For Halma, we handle defects as described in this paper, and for the superstabilizer approach, we chose~\cite{wei_low-overhead_2024} as the representative procedure to compare with. 

\vspace{2mm}
\textbf{Simulate Logical Error Rate:}

We perform circuit-level error simulation with a physical error rate of two-qubit gates ranging from $0.001$ to $0.01$. We employ the SI1000 noise model\cite{Gidney_2022_SI1000}. For each lattice, we run 1,000 shots of simulations to generate their error syndrome and use the open-source MWPM decoding provided by the PyMatching~\cite{higgott_sparse_2023} to calculate the logical error rate per round. For each distance, we group the lattices into 10 groups of equal size and calculate the mean logical error of each group. We then calculate the mean of the 10 groups with their standard deviation as uncertainty.
}

\vspace{2mm}

\begin{figure}[htbp]
\centering
\subfloat[Logical Error Rates]{\includegraphics[width=0.8\columnwidth, clip]{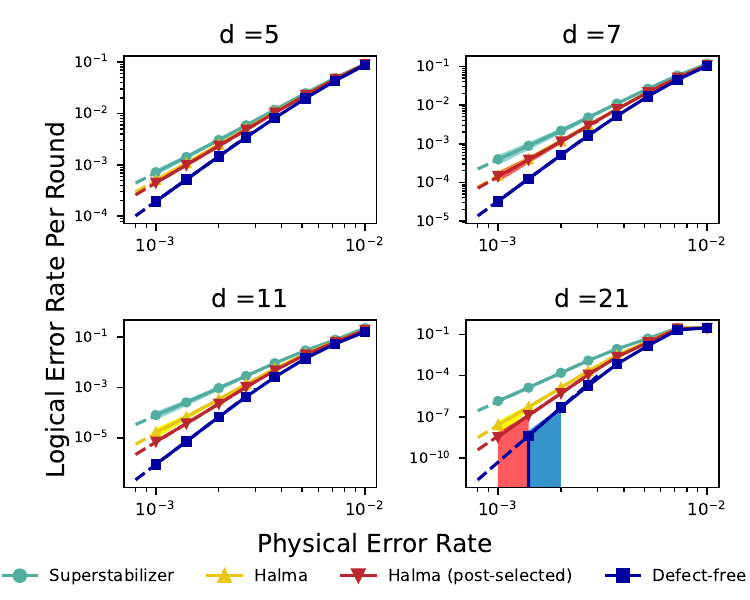}}\\
\subfloat[Teraquop Footprint]{\includegraphics[width=0.8\columnwidth, trim={0 0 0 33pt}, clip]{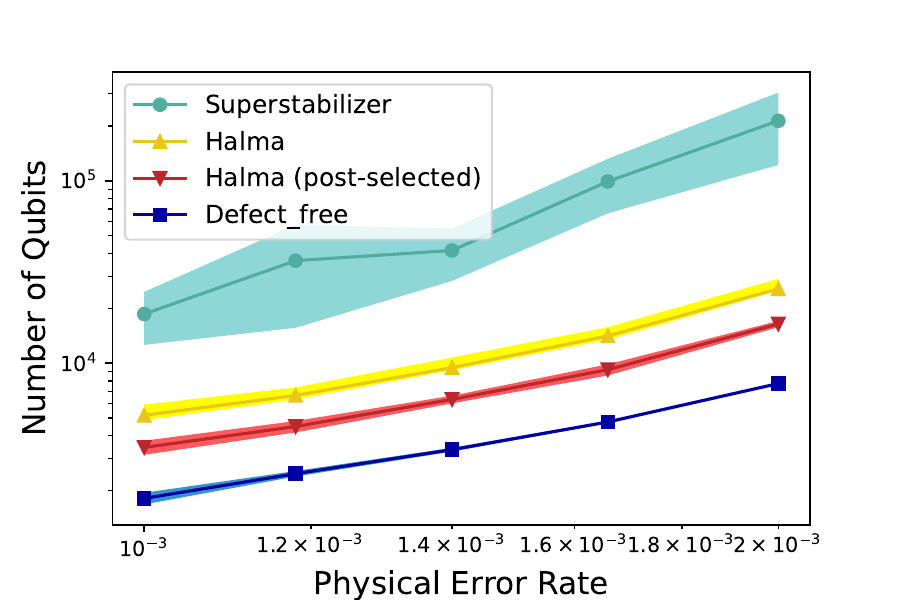} }
\caption{(a) Simulated Logical Error Rates (LER) per round for defect-free lattice (blue), defective lattice handled by Halma without post-selection (yellow), with post-selection (red), and superstabilizers (teal) with distances $d=5,7,11,21$. Vertical axes are the simulated Logical error rate per round, and horizontal axes are the physical error rate of two-qubit gates. (b) Teraquop footprint for surface code on defective lattices handled by superstabilizers, Halma with or without post-selection, and defect-free lattices. The code blocks are of size $d\times d \times d$ for the defect-free case, and $d\times d \times 2d$ for Halma and the superstabilizer approach, to compensate for the reduction in timelike distance.}
\label{MC_LER}
\end{figure}

We simulated the logical error rate for surface code with distance $d=5,7,9,11,13,15,21$. The resulting logical error rates of the simulations are displayed in Fig.~\ref{MC_LER}(a).

\begin{figure}[htbp]
\centering
\includegraphics[width=  0.6\columnwidth]{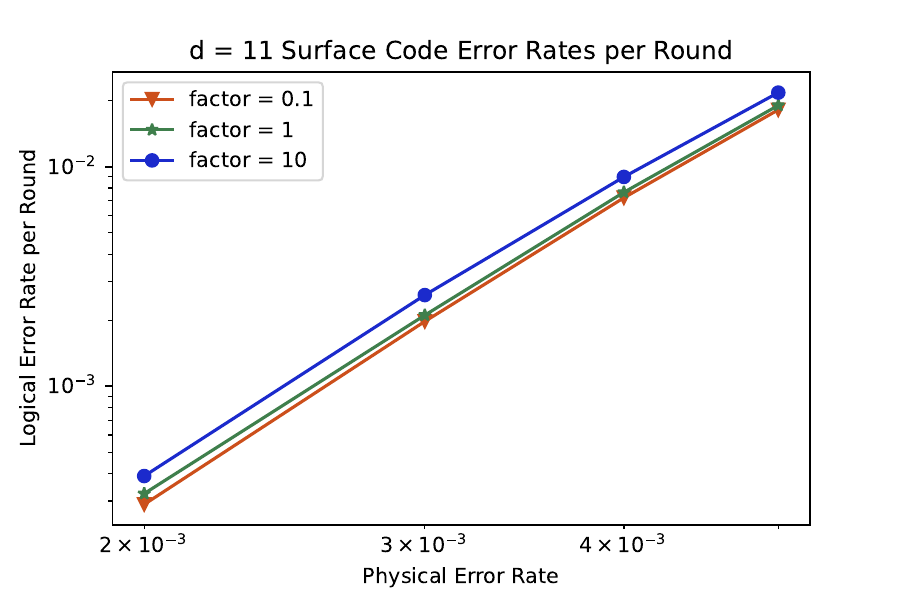}
\caption{The logical error rate averaged over 1000 randomly generated lattices with a defect rate of $2\%$ for a $d=11$ surface code, handled by Halma without post selection. The factor is the physical error rate of $\CXSWAP$ gates over that of the $\CNOT$ gates.}
\label{noise_factor}
\end{figure}

The SI1000 model assumes the gate noise of $\CNOT$ gates and $\CXSWAP$ gates to both be $p$. We also considered the effect of asymmetric physical error rates between the two types of two-qubit gates. Due to the infrequent appearance of $\CXSWAP$ gates in the syndrome extraction circuit, its physical error rate has a much less significant impact on the overall logical error rate. For a distance-11 surface code with a defect rate of 2\%, by increasing (decreasing) the physical error rate of $\CXSWAP$ gate by a factor of 10, the logical error rate of the code is only changed by less than $20\%$ ($10\%$), as shown in Fig.~\ref{noise_factor}.

For a certain physical error rate, we can fit the logical error to $p_{logic}= e^{ad+b}$, where $d$ is the defect-free distance of the code, $p_{logic}$ is the logical error rate per round, and $a, b$ are the parameters to be fitted. Using this fitted function, we can estimate the teraquop footprint of the surface code handled by different approaches, that is, the number of physical qubits needed to build a quantum memory with a logical error rate of $10^{-12}$. The result is shown in Fig.~\ref{MC_LER}(b).

We find that when handled by a conventional superstabilizer approach, the number of physical qubits needed is more than $10\times$ that of a defect-free lattice, with a physical error rate of $10^{-3}$. However, the qubit overhead is less than $3\times$ when handled by Halma, so it can reduce the qubit overhead by more than three times. This demonstrates the utility of Halma in the near-term realization of FTQC on hardware with fabrication defects.

With post-selection, the teraquop footprint becomes $\sim2\times$ that of a defect-free lattice, and the reduction in qubit overhead is further boosted to more than five times in comparison to the conventional superstabilizer approach. This stresses the potential improvement a sophisticated strategy module could provide.

\section{Discussion}
This paper introduced Halma, an ancilla-qubit-defect-mitigation technique based on an expanded set of native two-qubit gates. We showcased Halma's capability to perform cost-free qubit routing and mitigate ancilla qubit defects with no distance reduction. As we have demonstrated, the improvement provided by Halma will significantly ease the near-term realization of fault-tolerant quantum computing, in terms of the qubit overhead compensating for fabrication defects. 

Furthermore, we have recognized several directions that can be further studied to empower Halma to achieve its full potential, the first and foremost being an efficient strategy module that allows Halma to flexibly adapt to clusters of defects. As we have previously discussed, the simulation in this paper was done with either post-selection or a baseline strategy, which applies Halma only to isolated ancilla qubit defects. The design of a simple strategy module could follow the XXZZ 4-round cycle, and alternately handle conflicting ancilla qubit defects. This would apply Halma to a higher percentage of ancilla qubit defects, thus upholding a stronger preservation of the spacelike distance, in exchange of sacrificing the timelike distance of the code. One could also examine other heuristics, or attempt this problem with machine learning algorithms, to build a more sophisticated strategy module that could potentially have better performances. 

In addition, existing implementations of quantum error correction primarily rely on $\CNOT$-equivalent gates. Although the surface code is highly optimized for grid topology hardware, leaving limited room for improvement, alternative approaches, such as using $\iSWAP$ gates for syndrome extraction circuits~\cite{mcewen2023relaxing} still highlight the flexibility offered by expanded gate sets. Our work underscores that even for highly optimized surface code, leveraging a broader native gate set can effectively address the practical challenges of defects. Looking beyond surface codes, the expanded native gate set holds even greater promise, potentially enabling the implementation of some qLPDC codes on current hardware and reducing stringent hardware connectivity constraints.

\emph{Note Added.}---Upon finalizing our manuscript, we became aware of related work such as LUCI from Google~\cite{debroy_luci_2024} and even more recent SNL from AWS~\cite{leroux_snakes_2024}, both of which could mitigate ancilla qubit defects without reducing the code distance. We include a comparison with LUCI to provide readers with a clearer perspective on these works, while leaving the comparison with SNL for future studies to prevent any further delays. As we have previously discussed, in comparison to LUCI, Halma has the advantage of not further sacrificing the timelike distance of the code, thus enabling more efficient logical operations for small enough defect rates. Furthermore, Halma directly measures the values of the original stabilizers of surface code, instead of compiling superstabilizer (or stabilizer) values from the results of multiple gauge operator measurements. Therefore, it is less sensitive to measurement errors, which is the most significant source of error under the current noise model. This is a general advantage of Halma in comparison to most of the superstabilizer-based approaches, which typically rely on gauge operator measurements, including LUCI and SNL.

\section{Acknowledgement}
The work is supported by National Key Research and Development Program of China (Grant No. 2023YFA1009403), National Natural Science Foundation of China (Grant No. 12347104), Beijing Natural Science Foundation (Grant No. Z220002) and Zhongguancun Laboratory.

\section{Data Availability}
The data that support the findings of this article are openly available \cite{dataset}.

\nocite{dennis_topological_2002} 

\bibliography{Halma}

\end{document}